\documentclass[letter]{aa}    
\usepackage{float}
\usepackage{multirow}
\usepackage{makecell}
\usepackage{graphicx}
\usepackage{txfonts}
\usepackage{lipsum}

\usepackage{lscape}             
                                
\usepackage{placeins}           

\usepackage{graphicx, array, booktabs, tikz}
\usetikzlibrary{decorations.pathmorphing, arrows.meta, angles, quotes, calc}
\usepackage[colorlinks=true,allcolors=blue]{hyperref}
\usepackage{comment}

\begin{document}

\title{MeV absorption in gamma-ray bursts as a probe of their progenitor environments}

\author{
Gor~Oganesyan\inst{1,2,3}\thanks{\email{gor.oganesyan@gssi.it}}
\and
Om~Sharan~Salafia\inst{4,5}\thanks{\email{om.salafia@inaf.it}}
\and
Emanuele~Sobacchi\inst{1,2}\thanks{\email{emanuele.sobacchi@gssi.it}}
\and 
Samanta~Macera\inst{1,2}
\and 
Giancarlo~Ghirlanda\inst{4,5}
\and 
Lara Nava\inst{4,6}
\and 
Annarita Ierardi\inst{1,2}
\and 
Biswajit Banerjee\inst{1,2}
\and 
Alessio~Mei\inst{4}
\and 
Stefano Ascenzi\inst{1,2}
\and 
Marica~Branchesi\inst{1,2,3} 
}

\institute{
Gran Sasso Science Institute, Viale F. Crispi 7, L'Aquila (AQ), I-67100, Italy
\and 
INFN - Laboratori Nazionali del Gran Sasso, L'Aquila (AQ), I-67100, Italy
\and 
INAF - Osservatorio Astronomico d’Abruzzo, Via M. Maggini snc, I-64100 Teramo, Italy
\and 
INAF - Osservatorio Astronomico di Brera, via E. Bianchi 46, I23807 Merate (LC), Italy 
\and 
INFN – Sezione di Milano-Bicocca, Piazza della Scienza 3,
20146, Milano (MI), Italy
\and 
INFN, Sezione di Trieste, I-34127 Trieste, Italy
}

\date{Received xxx; accepted xxx}

\abstract{
A small fraction of X-ray photons from $\gamma$-ray bursts (GRBs), after escaping the relativistic jet, are scattered by electrons in the circumburst medium. Subsequent photon--photon absorption between the incoming MeV $\gamma$-rays and the back-scattered X-rays generate electron--positron pairs, enriching the surrounding medium with leptons. We investigate how these back-scattered photons modify the prompt GRB spectrum through $\gamma-\gamma$ absorption. In a dense and pair-loaded environment, the emerging spectra exhibit a broad absorption feature, whose profile is sensitive to the low-energy spectral index $\alpha$. In particular, spectra with $\alpha > -1$ develop a pronounced, saddle-shaped absorption between 1 and 100~MeV (rest frame). Such external MeV absorption could account for the spectral curvature seen in some bright GRBs, and may point to a dense circum-stellar medium (CSM) around their progenitor stars---consistent with early observations of core-collapse supernovae. In this scenario, the blastwave caused by the GRB is expected to start off with a relatively low Lorentz factor, and undergo an acceleration phase when traversing the large density drop at the interface between the dense CSM and the surrounding medium. The impact of these non-trivial dynamics on the afterglow emission is yet to be explored.
}


\maketitle
%
\section{Introduction}

Gamma-ray bursts (GRBs) are brief ($\sim$0.1--100~s) extragalactic transients releasing enormous energies of $10^{52}$--$10^{54}$~erg, predominantly observed in the keV--MeV range. A fraction of the prompt GRB photons, after escaping the relativistic jet where they are produced, are Thomson-scattered by cold electrons in the circumburst medium. The interaction between these back-scattered X-ray photons and the incoming MeV $\gamma$-rays leads to $\gamma$--$\gamma$ absorption and the subsequent creation of electron--positron pairs, which further enhance the scattering rate. Consequently, the immediate environment of the burst becomes enriched with pairs \citep{Thompson2000,Meszaros2001,Beloborodov2002}\footnote{The role of pair-enrichment by compact MeV sources has been first discussed in the context of Active Galactic Nuclei by \citet{Guilbert1983} and \citet{Beloborodov1999}.}. Continuous momentum deposition by the prompt emission photons accelerates this pair-enriched layer to ultra-relativistic velocities \citep{Madau2000,Beloborodov2002}. 

The influence of pair-loading on the early afterglow emission has been extensively studied \citep{Meszaros2001,Beloborodov2002,Kumar2004,Beloborodov2005,RR2007,Ghisellini2010,Nava2013}. However, its possible feedback on the prompt emission itself has received little attention. The enhanced lepton density could, in principle, cause significant absorption of MeV $\gamma$-rays by previously back-scattered X-ray photons via $\gamma$--$\gamma$ absorption. As noted by \citet{Meszaros2001} and \citet{Beloborodov2002}, such absorption features are expected to be most prominent in wind-like external media, where the high densities of free electrons at radii $\gtrsim10^{13}$--$10^{14}$~cm coincide with the region of prompt emission. 

In this Letter, we investigate spectral absorption features in GRB prompt emission induced by X-ray photons scattered by the circum-burst medium. We develop a semi-analytical model that self-consistently includes Thomson scattering, $\gamma$-$\gamma$ absorption, and the resulting pair-loading (Section~\ref{themodel}). We then apply this model to the prompt emission spectrum of GRB~190114C (Section~\ref{datacomparison}) to constrain the velocity of the wind and the mass-loss rate of its progenitor star. The broader implications of our results are discussed in Section~\ref{discussion}, followed by our conclusions in Section~\ref{conclusions}.

Hereafter, we adopt the notation $Q=10^{x}Q_{x}$, and cgs units unless otherwise specified, and a flat Cosmology with $\rm \Lambda CDM$ parameters from \citet{Planck2016}.

\section{The Model \label{themodel}}

\begin{figure*}
    \centering
    \includegraphics[width=0.80\textwidth]{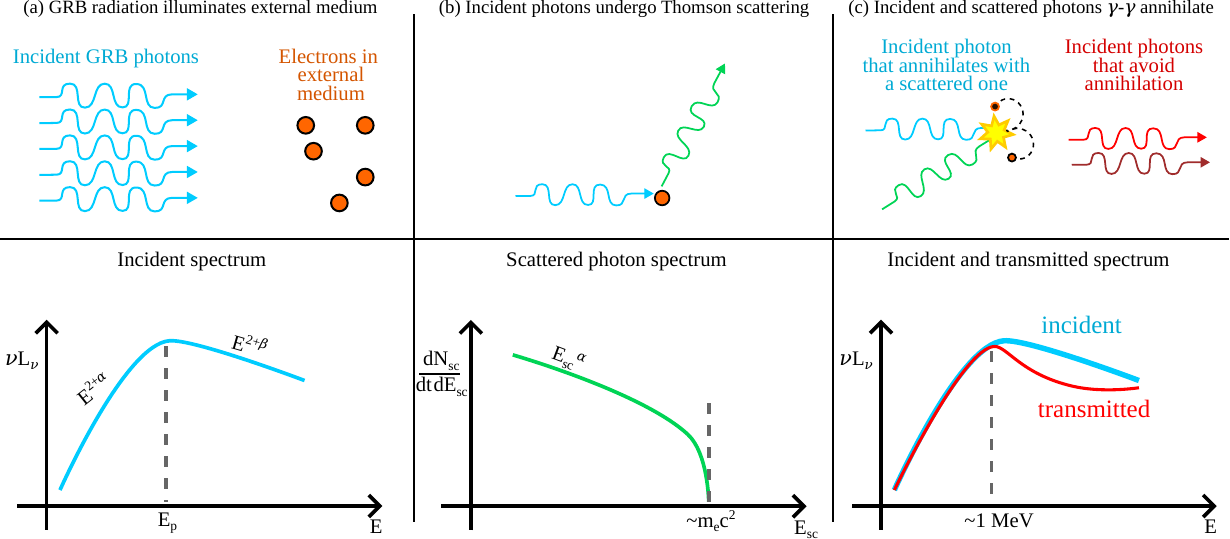}
    \caption{Sketch of the process that leads to the absorption feature that we discuss in this work. (a) The radiation from the GRB jet illuminates some external material located at a radius $R_0$. (b) Some of the photons in the incident GRB radiation undergo Thomson scattering off the electrons in the external medium. Because of the Klein-Nishina suppression of the Thomson cross section, the spectrum of the scattered photon spectrum cuts off at a photon energy $E_\mathrm{sc}\sim m_\mathrm{e}c^2$. (c) Incident photons above a threshold energy $E\gtrsim 2.5 m_\mathrm{e}c^2$ can annihilate with the scattered photons, leaving an absorption feature in the transmitted spectrum. If the low-energy photon index in the incident spectrum satisfies $\alpha>-1$, then the  feature is saddle-shaped.}
    \label{sketch}
\end{figure*}

The basic concept of external MeV absorption is illustrated in Fig.~\ref{sketch}. 
The GRB spectrum is assumed to be a peaked function in the $\nu L_{\nu}$ representation ($\mathrm{erg\,s^{-1}}$), with a low-energy slope corresponding to the single-electron synchrotron spectrum below the peak energy $E_{\rm p}$ (left-hand panels). 
Our choice of the low-energy slope is motivated by the findings of \citet{Oganesyan2017} and \citet[][see also \citealt{Toffano2021}]{Ravasio2018,Ravasio2019}, and we note that the median low-energy photon index of the brightest GRB spectra, modeled by the Band function, is -0.64 \citep{Gruber2014}.
A fraction of GRB photons undergo Thomson scattering on cold electrons in the circum-burst medium, producing a population of scattered photons (central panel). 
The rate of scattered photons per unit solid angle is determined by the differential Klein-Nishina cross section, with the characteristic energy $E_{\mathrm{max}}\approx 0.4\, m_{\rm e} c^2$. 
These scattered photons can then annihilate with the incoming MeV photons via $\gamma$-$\gamma$ pair production (\citealt{Madau2000}, right-hand panel). 

Because of the $\gamma$-$\gamma$ absorption threshold, this process becomes relevant only for incident photon energies above $E\approx 2.5\,m_{\rm e} c^{2}/(1+z)$ in the observer frame, where $z$ is the redshift of the source.
In the illustrative case considered here, with $\nu L_{\nu} \propto E^{2+\alpha}$ below the peak energy, and with $\alpha > -1$, the absorption weakens at the highest energies, producing a saddle-shaped absorption in the MeV range. 
This behavior arises because the number density of the scattered photons increases with energy. 
In contrast, spectra with $\alpha \le -1$ would exhibit a sharp suppression of all MeV photons above $2.5\,m_{\rm e} c^{2}/(1+z)$ (see Appendix \ref{sec:pair_enriched_medium}). 
Hence, the low-energy photon index $\alpha$ has a central role in determining the detailed shape of the MeV absorption feature. 

To quantify this picture, we consider a burst of un-polarised MeV radiation of a finite duration produced in a relativistic jet at a distance $R_{\gamma}$ from the central engine. 
We estimate the fate of the first photons escaping the jet and scattering off cold electrons in the circumburst medium. 
At a distance $R > R_{\gamma}$, an electron scatters the incoming photons at an average rate 
\begin{equation}
\frac{dN_{\mathrm{sc}}}{dt} \sim \frac{L\,\sigma_{\mathrm{T}}}{4\pi R^{2} m_e c^{2}},
\label{eq:rate}
\end{equation}
where $L$ is the isotropic-equivalent luminosity of the MeV front. 
The scattered photons have a typical survival time against $\gamma$-$\gamma$ absorption, as measured in the central engine rest frame, of 
\begin{equation}
\frac{\lambda_{\gamma\gamma}}{c} \sim \frac{4\pi R^{2} m_{\rm e} c^{2}}{\eta{\sigma}_{\rm T} L},
\label{eq:lambda_gg}
\end{equation}
where $\sigma_{\rm T}$ is the Thomson cross section, and $\eta\sigma_\mathrm{T}$ is the peak value of the $\gamma$-$\gamma$ cross section, with $\eta\sim 1/5$ \citep{Svensson1987}. 
For representative values of $R$ and $L$, one finds 
$\lambda_{\gamma\gamma}/c \sim 8\times10^{-4} R_{15}^{2} L_{53}^{-1} \,\mathrm{s}$ 
which, sufficiently close to the GRB emission site, is shorter than both the burst duration and its variability timescale. 

The optical depth to $\gamma$-$\gamma$ absorption for MeV photons that travel through the scattered X-ray photons can therefore be estimated as (see Appendix \ref{sec:abs_of_incident_photons} for a detailed treatment)
\begin{equation}
\tau_{\gamma\gamma} \sim Z_{\pm}\, \hat n(R_{0})R_{0} \eta{\sigma}_{\rm T} \frac{dN_{\mathrm{sc}}}{dt}\frac{\lambda_{\gamma\gamma}}{c}\sim Z_{\pm}\, \hat n(R_{0})R_{0} {\sigma}_{\rm T},
\end{equation}
where $\hat n(R_0)$ is the number density of the external medium, $Z_{\pm}$ is the lepton enhancement factor due to pair loading, and $R_{0}$ is the distance from the central engine where the dominant absorption takes place. This immediately shows that the absorption can be relevant (that is, $\tau_{\gamma\gamma}\gtrsim 1$) if the external density of leptons (after pair loading) is as high as
\begin{equation}
    Z_\pm\hat n(R_0)\gtrsim 1.5\times 10^9\,R_{0,15}^{-1}\,\mathrm{cm^{-3}}.
\end{equation}
If we assume a wind-like profile for the external medium, and we write $\hat n(R) = \hat A_\star (R/5.5\times 10^{17}\,\mathrm{cm})^{-2}$ \citep[so that $\hat A_\star=1\,\mathrm{cm^{-3}}$ corresponds to the density of a stellar wind with a mass loss rate $\dot M=10^{-5}\,\mathrm{M_\odot\,yr^{-1}}$ and a velocity $v_\mathrm{w}=10^3\,\mathrm{km\,s^{-1}}$, typical of a Wolf-Rayet star --][]{Chevalier1999}, then this translates to
\begin{equation}
    Z_\pm\hat A_\star\gtrsim 5\times 10^3\,R_{0,15}\,\mathrm{cm^{-3}}.
\end{equation}
In a wind-like circum-burst medium, most of the optical depth of this process is provided by the innermost layers, closest to $R_\gamma$. The scattering and pair enrichment of the medium, though, can accelerate these inner layers, inflating a cavity \citep{Madau2000,Beloborodov2002}: at a rest-frame time $t$ since the start of the prompt emission, the absorption is therefore mainly provided by the inner radius of the cavity, where the circum-burst medium has been loaded with pairs, but not yet accelerated. This corresponds to a radius \citep[][see also Appendix \ref{sec:pair_enriched_medium}]{Beloborodov2005}
\begin{equation}
    R_0(t)\sim 7.9\times 10^{15}\,L^{1/2}_{53}t^{1/2}_0 \xi^{-1/2}_\mathrm{acc,2}\,\mathrm{cm},
    \label{eq:R0}
\end{equation}
where $\xi_\mathrm{acc}\sim 100-150$ is a dimensionless parameter that depends on the shape of the incident spectrum, and $L$ here must be interpreted as the average luminosity until time $t$. At that radius, the pair-loading factor of the material that has not been accelerated yet by the prompt emission photons is $Z_\pm\sim 74$ \citep{Beloborodov2002}. The minimum required wind parameter for absorption effects to be relevant therefore becomes
\begin{equation}
    \hat A_\star \gtrsim 5.3\times 10^2 \,L^{1/2}_{53}t^{1/2}_0 \xi^{-1/2}_\mathrm{acc,2}\,\mathrm{cm^{-3}}.
\end{equation}

For soft GRB spectra ($\alpha < -1$; see Fig.~\ref{fig:example_abs_spectra}), the absorption produces a cut-off at high energy, which is visible already with $\hat{A}_\star \gtrsim \mathrm{few}\times 100\,\mathrm{cm^{-3}}$. For harder GRB spectra ($\alpha>-1$) and $\hat{A}_\star \gtrsim 1000\,\mathrm{cm^{-3}}$, a saddle-shaped absorption feature becomes visible.

At $R<R_0$, the medium is strongly accelerated. This blue-shifts the photon energy where $\gamma-\gamma$ absorption takes place, making it irrelevant for our purposes; in the accelerated medium, the Thomson opacity exceeds $\gamma-\gamma$ opacity \citep{Beloborodov2002}, but for our parameters of interest Thomson scattering does not modify appreciably the spectrum. Therefore, we only consider the effects of $\gamma-\gamma$ annihilation at $R\sim R_0(t)$, where the pair-loaded medium can produce an absorption feature at $\sim 1$--$10$ MeV. An accurate model that accounts for the detailed pair-loading profile is presented in Appendix~\ref{accuratemodel}.

\section{GRB 190114C as a test case \label{datacomparison}}
\begin{figure*}
\sidecaption
  \includegraphics[width=12cm]{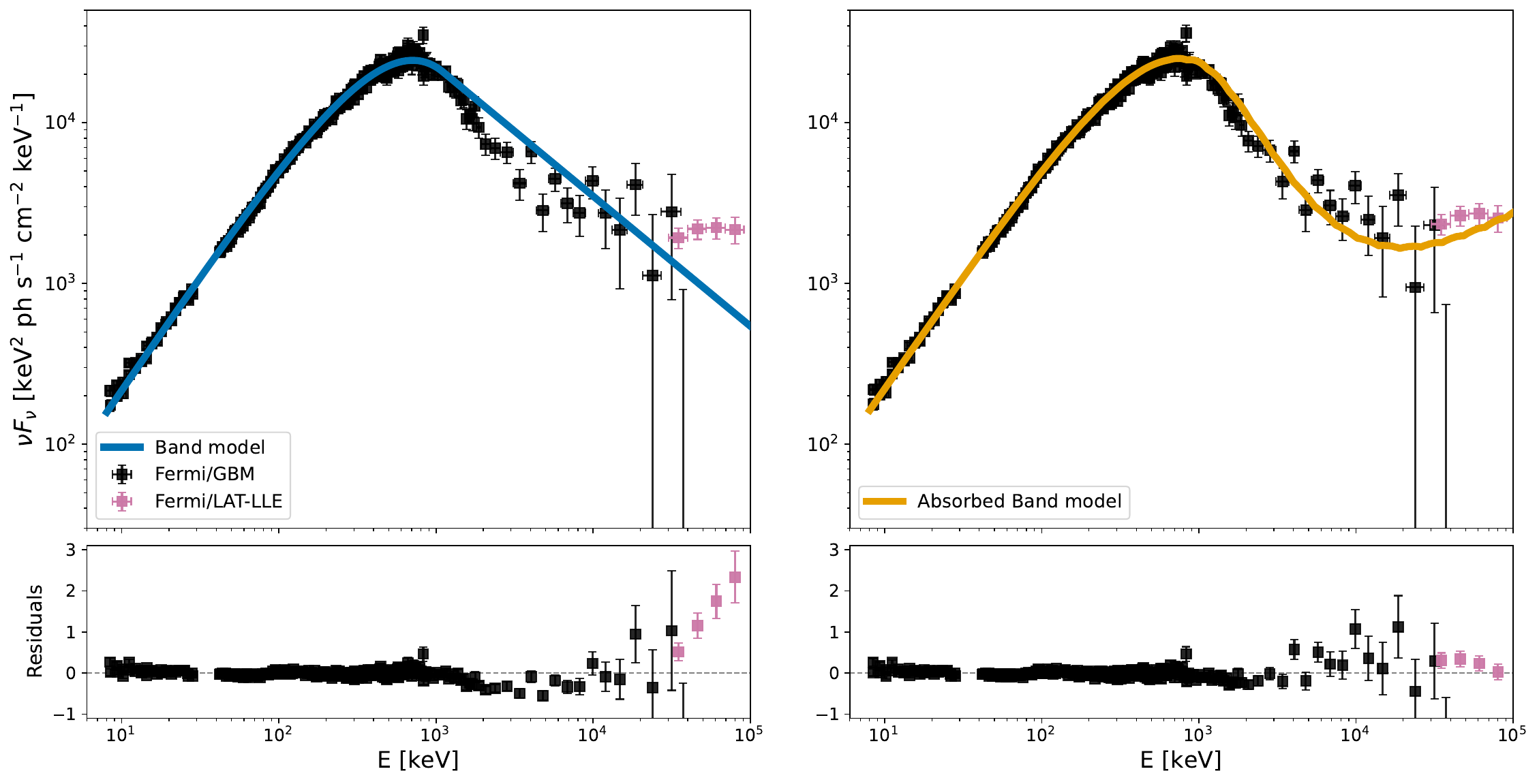}
     \caption{Comparison between the Band model (left-hand panel) and the absorbed Band model (right-hand panel) applied to the early prompt emission spectrum of GRB 190114C (0-4.15 s).\\[22pt]}
     \label{190114C_nuFnu}
\end{figure*}

Among the population of bright GRBs with well-sampled MeV coverage, we selected GRB~190114C, for which independent analyses have reported a sharp spectral cutoff around $\sim 1$~MeV, followed by a nearly flat spectrum up to $\sim 100$~MeV \citep{Ajello2020,Macera2025}. We analysed the early-time emission (0--4.15~s) of GRB~190114C by combining data from the \textit{Fermi} Gamma-ray Burst Monitor (GBM; 8~keV–40~MeV) and the Large Area Telescope Low-Energy (LAT-LLE) data set (30–100~MeV). The spectrum was initially fitted with the empirical Band function \citep{Band1993}, defined by four free parameters: the low- and high-energy photon indices $\alpha$ and $\beta$, the peak energy $E_{\mathrm{p}}$, and a normalisation constant. While the Band model reproduces the sub-MeV portion of the spectrum ($<1$~MeV) well, it fails to account for the pronounced curvature in the 1–100~MeV range (left panel of Fig.~\ref{190114C_nuFnu}). The residuals reveal a broad, saddle-shaped deviation across this energy band. 

To test the external absorption scenario, we implemented a custom model for the X-ray spectral fitting package \texttt{XSPEC} \citep{Arnaud1996}. The model has five free parameters: the low ($-1.5 < \alpha < -0.3$) and high-energy photon index ($-2.5 < \beta < -2.0$), the spectral peak ($2.5 < \log_{10}(E_{\mathrm{p}}/\mathrm{keV}) < 4.0$), the wind-density normalisation ($2 < \log_{10}(\hat{A}_\star/\mathrm{cm}^{-3}) < 5$), and the unabsorbed bolometric luminosity ($50 < \log_{10}({L}/\mathrm{erg \,} \mathrm{s}^{-1}) < 55$). The absorption radius $R_0$ is computed according to Eq.\ \ref{eq:R0}, where we experimented with different values of $\xi_\mathrm{acc}$ between 100 and 200 (to reflect the variation in this parameter for different spectral shapes and under different modeling assumptions; \citealt{Beloborodov2014}), and of $t_0$ between 1.5 and 3 s (to account for the fact that our model represents the instantaneous spectrum, but the observed spectrum is constructed from the counts accumulated between 0 and $4.15/(1+z)=2.9$ rest-frame seconds after the start of the GRB emission). The redshift of GRB~190114C was fixed at $z = 0.4245$ \citep{Castro-Tirado2019}.

Table~\ref{fit_results} lists the best-fit parameters for both the Band and absorbed Band models (the values in the table for the latter model correspond to $\xi_\mathrm{acc}=100$ and $t_0=2.9\,\mathrm{s}$), while Fig.~\ref{190114C_nuFnu} shows the corresponding $\nu F_{\nu}$ spectrum. Incorporating $\gamma$--$\gamma$ absorption improves the fit (right panel of Fig.~\ref{190114C_nuFnu}), with $\Delta \mathrm{stat} = 147$ for one additional parameter ($\hat{A}_\star$, independently on the chosen values of $\xi_\mathrm{acc}$ and $t_0$). This provides evidence that external absorption may play a significant role in shaping the prompt MeV spectrum of GRB~190114C. The best fit luminosity is $L\approx 3.5\times 10^{53}\,\mathrm{erg\,s^{-1}}$; when varying $\xi_\mathrm{acc}$ and $t_0$ in the mentioned ranges, the best-fit density \textbf{is} $1.7\times 10^3 < \hat A_\star/\mathrm{cm^{-3}} < 10^4\,$ and the absorption radius is $1.4 < R_0/10^{16}\,\mathrm{cm} < 2.5$.

\section{Discussion \label{discussion}}

\subsection{Implications for the progenitor medium}

The density profile of a stellar wind is set by the progenitor’s mass-loss rate $\dot{M}$ and wind velocity $v_{\rm w}$ prior to core collapse \citep{Chevalier1999}, $\hat n(R) = \dot{M}/4\pi m_{\rm p} v_{\rm w}R^2$, which defines the normalization $\hat A_\star = \dot M_{-5}v_{\mathrm{w},3}^{-1},\mathrm{cm^{-3}}$, with $\dot M$ expressed in $\mathrm{M_\odot ,yr^{-1}}$ and $v_\mathrm{w}$ in $\mathrm{km,s^{-1}}$. Long GRBs are commonly believed to be associated with Wolf–Rayet progenitors \citep{Woosley1993,Woosley2006}, whose mass-loss rates are uncertain, spanning $\dot{M} \sim 10^{-5}$–$10^{-3},\mathrm{M_{\odot},yr^{-1}}$ with velocities from a few hundred to several thousand $\mathrm{km,s^{-1}}$ \citep[e.g.][]{Vink2005,Vink2011,Sander2023,Pauli2025}.
However, there is growing evidence for significant circumstellar material (CSM) around some progenitors prior to explosion \citep[e.g.][]{Schlegel1990,Chugai2003,Chugai2004,Campana2006,Pastorello2007,Kiewe2012,Ofek2014,Groh2014,Khazov2016,Yaron2017,Salafia2026}. Such environments, also invoked for Type IIn supernovae (e.g. SN1994W; \citealt{Chugai2004}), deviate from a steady wind profile and may arise from eruptive mass ejection (e.g. luminous-blue-variable phases; \citealt{Kiewe2012}) or from binary evolution, during which a compact object embedded in a common envelope could expels large amounts of material \citep{Chevalier2012}.
If interpreted as an effective wind-like outflow, these scenarios imply extreme mass-loss rates up to $\dot{M} \sim 10^{-1}\mathrm{M_{\odot}},\mathrm{yr^{-1}}$, likely confined to the immediate vicinity of the progenitor, $R \approx 3\times10^{15},T_{\rm yr},v_{\rm w,3},\mathrm{cm}$, where $T_\mathrm{yr}$ is the time in years during which the outflow was produced prior to collapse. The parameters found in our study of GRB190114C correspond to $\dot M \sim 10^{-2}-10^{-1}\,\mathrm{M_\odot\,yr^{-1}}$ and $T_\mathrm{yr}\sim \mathrm{3-10}$ and are thus consistent with these expectations.

The presence of a dense CSM around the progenitor can have an important impact on the dynamics and emission of the external shock that arises from the interaction of the GRB ejecta with the external medium. In Appendix \ref{app:denseafterglow} we discuss briefly this aspect and touch on the compatibility of the proposed scenario with the observed afterglow, especially for what concerns the very high energy emission seen by MAGIC \citep{MAGIC1-2019}. While we did not find obvious contradictions, we note that further investigations are needed in order to conclude whether the available information on GRB190114C is fully consistent with our proposed scenario.

\subsection{Relevance to the observed prompt emission features}

If the wind density requirements are met, the spectra of collapse-driven GRBs are expected to exhibit absorption of MeV $\gamma$-rays above $2.5\,m_{\rm e}c^{2}$ in the rest frame. The detailed shape of this absorption feature -- whether a saddle-like or an exponential cutoff -- depends on the low-energy slope of the MeV pulse and on the position of the characteristic energy. Several observed spectral features could naturally arise in this scenario.  

First, the characteristic spectral peak of GRBs rarely exceeds $\sim1$\,MeV\citep{Gruber2014}\footnote{The exact value of the spectral peak depends on the spectral model adopted in the fit.}. Second, the MeV absorption tends to sharpen the overall width of the observed spectrum, potentially alleviating the long-standing difficulty of reproducing GRB spectra with simple non-thermal emission models.  

Interestingly, the fixed threshold for MeV absorption, corresponding to $\sim 2.5\,m_{\rm e}c^{2}$ in the rest frame, implies a strong redshift dependence of the observed spectral shape. A high-redshift GRB would therefore display the MeV dip or cutoff feature shifted into the hard X-ray band ($\gtrsim 100$\,keV). Then, in principle, the identification of this absorption feature can be used as a redshift indicator.
  
\section{Conclusions \label{conclusions}}

We have presented a simple model describing the absorption of MeV $\gamma$-rays caused by backscattering of GRB photons in a cold circum-burst medium. When applied to the prompt emission spectrum of GRB~190114C, the model provides a significantly better fit than the standard Band function. 

A noticeable MeV absorption requires a dense circum-stellar material around the progenitor. The detailed shape of the MeV absorption feature depends sensitively on the low-energy spectral slope of the MeV pulse, while the absorption threshold at $\approx 2.5\,m_{e}c^{2}$ (rest frame) introduces a strong redshift dependence in the observed spectra.  

This mechanism could account for some common properties of GRB spectra: (i) the characteristic spectral peaks rarely exceed $\sim1$\,MeV in the rest frame, and (ii) the relatively sharp spectral shapes near the peak energy, which are sometimes difficult to reproduce with standard non-thermal emission models in optically thin jet regions.

\begin{acknowledgements}
We thank the anonymous referee for the comments that have improved this letter. We thank P. Blasi, E. Kammoun, A. Romagnolo, S. Boula, F. Aharonian and E. Amato for fruitful discussions. This research has made use of data obtained through the High Energy Astrophysics Science Archive Research Center Online Service, provided by the NASA/Goddard Space Flight Center, and specifically, this work made use of public Fermi-GBM and Fermi-LAT data. B.B. acknowledges financial support from the Italian Ministry of University and Research (MUR) for the PRIN grant METE under contract no. 2020KB33TP.  M.B. and G.O. acknowledge the ACME project, which has received funding from the European Union’s Horizon Europe Research and Innovation program under Grant Agreement No. 101131928. This work has been funded by the European Union-Next Generation EU, PRIN 2022 RFF M4C21.1 (202298J7KT - PEACE). O.S.S. acknowledges funding from INAF through grant 1.05.23.04.04.
\end{acknowledgements}

\bibliographystyle{aa}
\bibliography{bibliography}

\begin{appendix}
\onecolumn
\section{The MeV absorption model \label{accuratemodel}}

Here we provide the details of our model for $\gamma$–$\gamma$ absorption of MeV photons. We assume that the photons have already escaped the jet and neglect its finite angular size, treating the photon injection as one-directional.

\subsection{Single electron in the field of incident photons}

A single cold electron (or positron) in an un-perturbed external medium scatters the incoming photons at an average rate (per unit scattered photon energy $\varepsilon_\mathrm{sc}$ -- in units of $m_\mathrm{e} c^{2}$ -- and unit solid angle at a scattering cosine angle $\mu=\cos\theta$)  
\begin{equation}
    \frac{dN_{\rm sc}}{dt\,d\varepsilon_\mathrm{sc}\,d\Omega}(\varepsilon_\mathrm{sc},\mu) = \int \frac{d\sigma_{\text{KN}}}{d\Omega}(\varepsilon, \mu) \, \frac{dn_{\gamma}}{d\varepsilon} \, c \, \delta(\varepsilon_\mathrm{sc} - \varepsilon/(1+(1-\mu)\varepsilon)) \, d\varepsilon, 
\end{equation}
where $\varepsilon$ is the energy of the incident photon, $d\sigma_{\text{KN}}/d\Omega$ is the differential Klein-Nishina cross-section, and 
\begin{equation}
    \frac{dn_{\gamma}}{d\varepsilon} = \frac{ L_{\varepsilon} }{4\pi R^2 m_{\rm e} c^3 \varepsilon}
\end{equation}
is the specific number density of the incident photons. Here $R$ represents the distance from the central engine where the scattering takes place, $L_{\varepsilon}=dL/d\varepsilon$ the specific luminosity of the incident photons, and $\varepsilon/(1+(1-\mu)\varepsilon$
the energy of the scattered photon. 
A simple transformation of the Dirac delta function 
\begin{equation}
    \delta(\varepsilon_\mathrm{sc} - \varepsilon/(1+(1-\mu)\varepsilon)) = \frac{\delta(\varepsilon - \varepsilon_0)}{\left(1-(1-\mu)\varepsilon_\mathrm{sc} \right)^{2}},
\end{equation}
where  
\begin{equation}
    \varepsilon_{0} = \frac{{\varepsilon_\mathrm{sc}}}{1-(1-\mu){\varepsilon_\mathrm{sc}}},
\end{equation}
leads to the final expression for the single-electron scattering rate as a function of the incident spectrum,
\begin{equation}
\frac{dN_{\rm sc}}{dt\, d\varepsilon_\mathrm{sc}\, d\Omega} = \frac{d\sigma_{\mathrm{KN}}}{d\Omega}(\varepsilon_0, \mu) \cdot \frac{L_{\varepsilon}(\varepsilon_0)}{4\pi R^2 m_{\rm e} c^2} \cdot \frac{1}{\varepsilon_\mathrm{sc} \left[1 - (1 - \mu)\varepsilon_\mathrm{sc}\right]}.
\end{equation}

\subsection{Propagation of the scattered photons}

A scattered photon of energy $\varepsilon_\mathrm{sc}$ and scattering angle $\mu$ can undergo $\gamma-\gamma$ absorption when interacting with another incident photon. The mean free path of the scattered photon due to such absorption process can be defined as  
\begin{equation}
    \lambda^{-1}_{\gamma \gamma}(\varepsilon_\mathrm{sc},\mu) = \int_{\varepsilon_{\rm th}}^{+\infty} (1-\mu) \sigma_{\gamma \gamma}(\varepsilon_\mathrm{sc},\varepsilon,\mu) \frac{dn_{\gamma}}{d\varepsilon} d\varepsilon,
    \label{lmbd_exact_solution}
\end{equation}
where $\varepsilon$ represents again the energy of an incident photon, $\varepsilon_{\rm th}=2/(1-\mu)\varepsilon_\mathrm{sc}$ is the threshold for $\gamma-\gamma$ absorption, and 
\begin{equation}
\sigma_{\gamma \gamma}(\varepsilon_\mathrm{sc},\varepsilon,\mu)=\frac{3\sigma_\mathrm{T}}{8x^2}\left[\left(2+2x^{-2}-x^{-4}\right)\ln\left(x+\sqrt{x^{2}-1}\right)
 -\left(1+x^{-2}\right)\sqrt{1-x^{-2}}\right],
\end{equation}
is the Breit-Wheeler cross section, with $x=\sqrt{\varepsilon_\mathrm{sc}\varepsilon(1-\mu)/2}$ \citep{Jauch1976}. 

\begin{figure*}
\sidecaption
  \includegraphics[width=8cm]{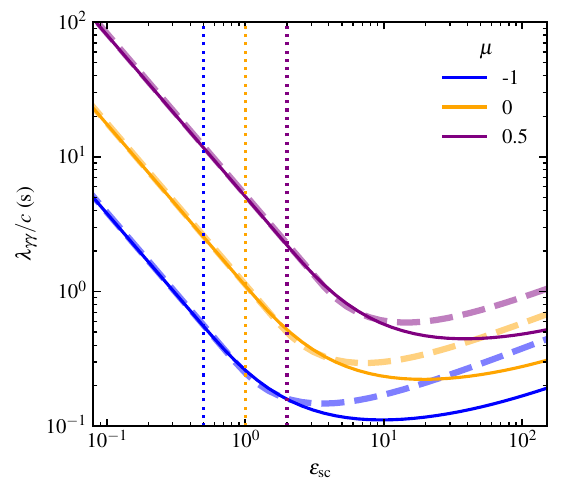}
     \caption{Exact and approximate $\gamma-\gamma$ absorption mean free path. Solid lines show the exact $\gamma-\gamma$ absorption mean free path (Eq.\ \ref{lmbd_exact_solution}) of a photon with energy $\varepsilon_\mathrm{sc}$ scattered at a radius $R_0=10^{16}\,\mathrm{cm}$ to a cosine angle $\mu$ (different colors refer to different values of $\mu$ as given in the legend) moving through an incident radiation with $L=10^{53}\,\mathrm{erg/s}$, $E_\mathrm{p}=1\,\mathrm{MeV}$, $\alpha=-0.6$ and $\beta=-2.2$. Dashed lines show the corresponding approximate mean free path (Eq.\ \ref{lmbd_appr_solution}) setting $\eta=0.35$. Each dotted line shows the maximum energy of a scattered photon whose
    scattering angle $\mu$ is given by the corresponding color in the legend.\\[22pt] }
     \label{fig:lmbd_comparison}
\end{figure*}

A good approximation of $\lambda_{\gamma \gamma}$ can be obtained by approximating $\sigma_{\gamma\gamma}$ as a delta function centered at $\varepsilon=2\varepsilon_\mathrm{th}=4/(1-\mu) \varepsilon_\mathrm{sc}$ (i.e.\ where the cross section is maximised), with a normalization $\sigma_{\gamma\gamma}\sim \eta \sigma_\mathrm{T} \delta(\varepsilon-2\varepsilon_\mathrm{th})$ that is set by a constant $\eta$ \citep[e.g.][]{Svensson1987}. This results in
\begin{equation}
    \lambda^{-1}_{\gamma \gamma}(\varepsilon_\mathrm{sc},\mu) \sim \frac{L_{\varepsilon}\left(\frac{4}{(1-\mu)\varepsilon_\mathrm{sc}}\right) \eta \sigma_{\rm T}(1-\mu)}{4\pi R^{2} m_{\rm e} c^3},
    \label{lmbd_appr_solution}
\end{equation}
As shown in Fig.\ref{fig:lmbd_comparison}, for our assumed incident spectrum this approximation is quite accurate when setting $\eta\approx 0.35$. In our implementation, we make use of this approximation when computing $\lambda_{\gamma\gamma}$ for the scattered photons. Conversely, we use the exact formula when computing the $\gamma-\gamma$ optical depth for incident photons, because this affects the detailed shape of the absorbed spectrum.

\subsection{Absorption of incident photons}
\label{sec:abs_of_incident_photons}
Let us now indicate with $t=0$ the time when the first incident photon reaches the radius $R$ of interest. At a later time $t$, the average number density of scattered photons (per unit solid angle and per unit energy of the scattered photon $\varepsilon_\mathrm{sc}$) is obtained\footnote{Because all the process happens in a time less than $r/c$, scattered photons can be assumed to remain close to their scattering centres.} by summing over all photons scattered at times $0<t_\mathrm{sc}<t$, and accounting for the fact that only a fraction $e^{-c(t-t_{\rm sc})/\lambda_{\gamma\gamma}}$ has survived to $\gamma-\gamma$ absorption, namely 
\begin{equation}
    \frac{dn_{\gamma}}{d\varepsilon_\mathrm{sc} d\Omega} (R,\varepsilon_\mathrm{sc},\mu,t) \sim \int_{0}^{t} \hat{n}(R,t_{\rm sc}) \frac{dN_{\rm sc}}{dt d\varepsilon_\mathrm{sc} d\Omega}(\varepsilon_\mathrm{sc},\mu) e^{-\frac{c(t-t_{\rm sc})}{\lambda_{\gamma \gamma}(\varepsilon_\mathrm{sc},\mu)}}dt_{\rm sc}.
    \label{eq:scattered_photon_density}
\end{equation}
Here $\hat{n}(R,t_{\rm sc})$ is the lepton number density of the medium at time $t_{\rm sc}$. This includes both the original electrons and the pairs created during the process. Under the assumption that the incident spectrum is constant, and in a case where the lepton number density remains constant with time (so that $\hat{n}(R,t_\mathrm{sc})\sim \hat{n}(R,0)=\hat{n}(R)$, i.e.\ the pair enrichment is negligible) we can simplify the above expression as 
\begin{equation}
    \frac{dn_{\gamma}}{d\varepsilon_\mathrm{sc} d\Omega} (R,\varepsilon_\mathrm{sc},\mu,t) \sim  \hat{n}(R) \frac{dN_s}{dt d\varepsilon_\mathrm{sc} d\Omega}(\varepsilon_\mathrm{sc},\mu) \frac{\lambda_{\gamma\gamma}(\varepsilon_\mathrm{sc},\mu)}{c} \delta \hat{t} (\varepsilon_\mathrm{sc},\mu,t), 
\end{equation}
where 
\begin{equation}\label{t_hat}
    \delta \hat{t} (\varepsilon_\mathrm{sc},\mu,t) =  1 - e^{-\frac{ct}{\lambda_{\gamma \gamma}(\varepsilon_\mathrm{sc},\mu)}}.
\end{equation}

\begin{figure*}
  \centering
  \sidecaption
  \includegraphics[width=8cm]{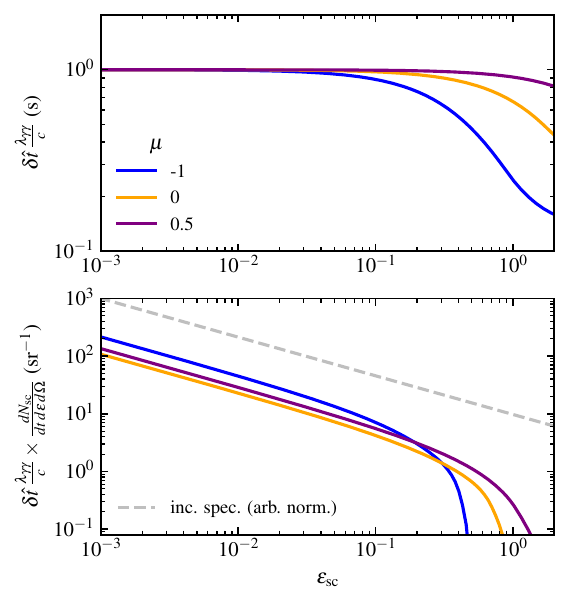}
  \caption{Scattered photon survival time and corresponding spectrum. We have assumed $t_0=1$ s, $\alpha=-2/3$, $\beta=-2.2$, $E_{p}=1$ MeV, $L=10^{53}$ erg/s and $R=10^{16}\,\mathrm{cm}$. The solid lines in the top panel show the survival time of scattered photons with three different scattering angles (given in the legend) before they completely annihilate with incident photons. Lines with the same colors in the bottom panel show the corresponding spectrum of these scattered photons, expressed as the product of the survival time and the specific scattering rate. The grey dashed line shows the incident spectrum, arbitrarily normalized, for comparison.\\[22pt]}
  \label{fig:dt_scattered}
\end{figure*}

\noindent This shows that, at time $t$, the photon density is essentially set by photons scattered during a time interval preceding $t$ of duration $\lambda_{\gamma\gamma}\delta \hat t(\varepsilon_\mathrm{sc},\mu)/c\leq \lambda_{\gamma\gamma}/c$ , because photons scattered at earlier times have entirely annihilated by then. We call this the `survival time' of scattered photons.  
In Fig.~\ref{fig:dt_scattered} we show examples of $\delta \hat{t}$ and the corresponding single-scattering spectrum.

The above result also implies that, even in a case where the lepton density $\hat n$ evolves over time (because of pair enrichment), the scattered photon density only depends on an appropriately averaged density of the leptons over the latest $\lambda_{\gamma\gamma}/c$ time. Alternatively, the evolution of $\hat n$ can be seen as an additional factor that modifies the survival time $\delta \hat t$ into a new, effective one $\delta \hat t_\mathrm{eff}$. This can be seen by writing the lepton density in Eq.\ \ref{eq:scattered_photon_density} as $\hat n(R,t_\mathrm{sc}) = \hat n(R,0)\times(\hat n(R,t_\mathrm{sc})/\hat n(R,0))$, which leads to
\begin{equation}
    \frac{dn_{\gamma}}{d\varepsilon_\mathrm{sc} d\Omega} (R,\varepsilon_\mathrm{sc},\mu,t) \approx \hat{n}(R,0)\frac{dN_{\rm sc}}{dt d\varepsilon_\mathrm{sc} d\Omega}(\varepsilon_\mathrm{sc},\mu)\int_{0}^{t} \frac{\hat n(R,t_\mathrm{sc})}{\hat n(R,0)}  e^{-\frac{c(t-t_{\rm sc})}{\lambda_{\gamma \gamma}(\varepsilon_\mathrm{sc},\mu)}}dt_{\rm sc} \equiv \hat{n}(R,0)\frac{dN_{\rm sc}}{dt d\varepsilon_\mathrm{sc} d\Omega}(\varepsilon_\mathrm{sc},\mu) \frac{\lambda_{\gamma\gamma}(\varepsilon_\mathrm{sc},\mu)}{c}\delta \hat t_\mathrm{eff}(\varepsilon_\mathrm{sc},\mu),
    \label{eq:scattered_photon_density_varying_n}
\end{equation}
with
\begin{equation}
    \delta \hat t_\mathrm{eff}(\varepsilon_\mathrm{sc},\mu) = \int_{0}^{t} \frac{\hat n(R,t)}{\hat n(R,0)}  e^{-\frac{c(t-t_{\rm sc})}{\lambda_{\gamma \gamma}(\varepsilon_\mathrm{sc},\mu)}}d\frac{c t_{\rm sc}}{\lambda_{\gamma\gamma}}=e^{-ct/\lambda_{\gamma \gamma}}\int_{0}^{ct/\lambda_{\gamma\gamma}} \frac{\hat n(R,x\lambda_{\gamma\gamma}/c)}{\hat n(R,0)}  e^{x}dx.
    \label{eq:that_eff}
\end{equation}
It is straightforward to see that the above equation reduces to Eq.\ \ref{t_hat} in the case of a constant lepton density.

The differential optical depth to $\gamma-\gamma$ absorption for an incident photon travelling at time $t$ through the field of photons previously scattered by material located within $(R,R+dR)$ is 
\begin{equation}\label{dtau_dR}
\frac{d\tau_{\gamma \gamma}}{dR}(\varepsilon,R,t) = 2\pi \int_{-1}^{1} \int_{\varepsilon_{\rm th,sc}}^{+\infty} \frac{dn_\gamma}{ d\varepsilon_\mathrm{sc}\, d\Omega} (\varepsilon_\mathrm{sc},\mu,t)  \, (1-\mu) \,\sigma_{\gamma \gamma}(\varepsilon,\varepsilon_\mathrm{sc},\mu) d\varepsilon_\mathrm{sc} d\mu = 2\pi \int_{-1}^{1} \int_{\varepsilon_{\rm th,sc}}^{+\infty} \frac{dn_\gamma}{ d\varepsilon_\mathrm{sc}\, d\Omega} (\varepsilon_\mathrm{sc},\mu,t)  \, (1-\mu) \,\sigma_{\gamma \gamma}(\varepsilon,\varepsilon_\mathrm{sc},\mu) d\varepsilon_\mathrm{sc} d\mu, 
\end{equation}
where $\varepsilon_{\rm th,sc} = 2/(1-\mu)\varepsilon$ is the threshold scattered photon energy for it to cause $\gamma-\gamma$ absorption of the incident photon.  
As discussed in the main text, in our scenario most of the optical depth at time $t$ is provided by scattering close to a specific radius $R=R_0(t)$. Under the assumption of a wind-like external profile, so that $\hat n(R,0)= \hat A R^{-2}$, and using approximation \ref{lmbd_appr_solution} for $\lambda_{\gamma\gamma}$, we can then write the total optical depth as 
\begin{equation}\label{tau_gg}
\tau_{\gamma \gamma} (\varepsilon,t) \sim R_0 \frac{d\tau_{\gamma\gamma}}{dR}(R_0) \sim  \frac{2 \pi \hat{A}}{R_{0}} \int_{-1}^{1}  \int_{\frac{2}{(1-\mu)\varepsilon}}^{+\infty} \frac{d\sigma_{\mathrm{KN}}}{d\Omega}({\varepsilon}_0, \mu) \frac{L_{{\varepsilon}}({\varepsilon}_0)}{ L_{{\varepsilon} }({2\varepsilon}_{\rm th})} \frac{\sigma_{\gamma \gamma}(\varepsilon,\varepsilon_\mathrm{sc},\mu)}{\eta \sigma_T} \frac{\delta \hat t_\mathrm{eff}(\varepsilon_{\rm sc},\mu)}{\varepsilon_\mathrm{sc} \left[1 - (1 - \mu)\varepsilon_\mathrm{sc}\right]} d\varepsilon_\mathrm{sc} d\mu, 
\end{equation}
where again $\varepsilon_0=\varepsilon_\mathrm{sc}/(1-(1-\mu)\varepsilon_\mathrm{sc})$ and $2{\varepsilon}_{\rm th}=4/(1-\mu)\varepsilon_\mathrm{sc}$.  

For a very high column density $\hat{A}/R_{0}$, Thompson scattering can further affect the low-energy spectrum, thus the overall optical depth is $\tau ({\varepsilon}) \approx \tau_{\rm \gamma e}({\varepsilon})+\tau_{\gamma \gamma}({\varepsilon})$ where $\tau_{\rm\gamma e}({\varepsilon}) \sim  \frac{\hat{A}}{R_{0}} \sigma_{\rm{KN}}({\varepsilon})$ and $\sigma_{\rm{KN}}({\varepsilon})$ is the total Klein-Nishina cross section. 

\subsection{Pair-enriched medium}
\label{sec:pair_enriched_medium}

The electron-positron pairs produced by the annihilation of the scattered X-ray radiation amplifies the density of leptons \citep{Madau2000}. Before the deposition of momentum into the outer medium by scattering and pair production becomes significant, the lepton density amplification factor can be written as \citep{Beloborodov2002}
\begin{equation}\label{loading}
    \frac{\hat n(R,t_\mathrm{sc}) }{\hat n (R,0)} = \frac{1}{2} (e^{\xi(R,t_\mathrm{sc})/\xi_{\rm load}}+e^{-\xi(R,t_\mathrm{sc})/\xi_{\rm load}}), 
\end{equation}
where $\xi(R,t_\mathrm{sc}) = c t_\mathrm{sc}/\lambda_{\rm e\gamma}$ is a dimensionless time coordinate, $\lambda_{\rm e\gamma} \approx 4 \pi R^{2} m_{\rm e} c^3/L \sigma_{\rm T}$ is the mean free path of an electron in the field of incident photons, and $\xi_{\rm load}(E_\mathrm{p},\alpha,\beta) \sim 20-30$ is the characteristic pair-enrichment scale which depends on the spectral shape (set by $E_\mathrm{p}$, $\alpha$ and $\beta$) of the incident photons\footnote{\citet{Beloborodov2005} derived a formula for $\xi_\mathrm{load}$ that depends on $\alpha$ and $\beta$ only, assuming that $E_\mathrm{p}=511$ keV, while no general formula exists yet to our knowledge. 
}. The exponential growth of pairs effectively continues until $\xi_{\rm acc} \approx 5 \xi_{\rm load}$ (assuming an external medium composed primarily of hydrogen), after which the medium accelerates to ultra-relativistic velocities. Once the pair-enriched layer of the medium is ultra-relativistic, the pair-loading gets suppressed and the MeV absorption rapidly shifts to higher photon energies. The absorption effects discussed in this work are therefore relevant at scattering times $t_\mathrm{sc}\leq t_{\rm acc}$, where $t_{\rm acc} = \xi_{\rm acc} \lambda_{\rm e \gamma}/c$. At later times, for simplicity, we approximate the effect of the radiative acceleration of the external medium simply as a cut-off in the external density, $\hat n(R,t_\mathrm{sc}>t_\mathrm{acc})=0$, but we stress that these times are not of interest for our investigation: when radiative acceleration is important at a radius $R$ (and hence absorption is suppressed), then the relevant radius for absorption will shift to a larger radius $R_0>R$ for which the acceleration has not become effective yet. Such radius can be estimated by setting $\xi_\mathrm{acc}=c t/\lambda_\mathrm{e\gamma}$, which yields
\begin{equation}
    R_0(t) \sim \left(\frac{\sigma_\mathrm{T}Lt}{4\pi m_\mathrm{e}c^2\xi_\mathrm{acc}}\right)^{1/2}\approx 7.9\times 10^{15}\,L^{1/2}_{53}t^{1/2}_0 \xi^{-1/2}_\mathrm{acc,2}\,\mathrm{cm}.
\end{equation}
For a given choice of $t$, radii $R\ll R_0$ do not contribute significantly to the absorption because they have been radiatively accelerated; larger radii $R\gg R_0$ do not contribute significantly because they are less dense (assuming a wind-like density profile) and because they are less pair-enriched. Hence, most of the absorption happens around $R=R_0(t)$.  

That said, the effective time $\delta\hat{t}_\mathrm{eff}(\varepsilon_\mathrm{sc},\mu)$ in the pair-enriched medium case, from Eq.\ \ref{eq:that_eff}, is 
\begin{equation}
    \delta\hat{t}_{\rm eff}(\varepsilon_\mathrm{sc},\mu,t) = \frac{1}{2} \int_{0}^{\min(t,t_{\rm acc})} \left(e^{\frac{ct_{\rm sc}}{\lambda_{\rm e\gamma} \xi_{\rm load}}}+e^{-\frac{ct_{\rm sc}}{\lambda_{\rm e\gamma} \xi_{\rm load}}} \right) e^{-\frac{c}{\lambda_{\gamma\gamma}(\varepsilon,\mu)}(t-t_{\rm sc})}\frac{c}{\lambda_{\gamma\gamma}}dt_{\rm sc}=\frac{e^{-ct/\lambda_{\gamma \gamma}}}{2}\int_{0}^{\min(t,t_\mathrm{acc})c/\lambda_{\gamma\gamma}} \left(e^{\frac{\lambda_{\gamma\gamma}}{\lambda_{\rm e\gamma} \xi_{\rm load}}x}+e^{-\frac{\lambda_{\gamma\gamma}}{\lambda_{\rm e\gamma} \xi_{\rm load}}x} \right) e^{x}dx.
\end{equation}
The integral is analytical and it formally results in 

\begin{equation}\label{t_hat_pairs}
    \delta \hat{t}_{\rm eff}(\varepsilon_\mathrm{sc},\mu,t) = \frac{e^{-ct/\lambda_{\gamma\gamma}}}{2} \left(\frac{e^{(1+\rho_\mathrm{acc}) c\min(t,t_\mathrm{acc})/\lambda_{\gamma\gamma}}}{1+\rho_\mathrm{acc}}+\frac{e^{(1-\rho_\mathrm{acc}) c\min(t,t_\mathrm{acc})/\lambda_{\gamma\gamma}}}{1-\rho_\mathrm{acc}}-\frac{2}{1-\rho_\mathrm{acc}^2}\right) , 
\end{equation}
where 
\begin{equation}
\rho_\mathrm{acc}(\varepsilon_\mathrm{sc},\mu)=\frac{\lambda_{\gamma\gamma}(\varepsilon_\mathrm{sc},\mu)}{\xi_\mathrm{load}\lambda_\mathrm{e\gamma}} \approx \frac{L}{\xi_\mathrm{load}L_{\varepsilon}\left(\frac{4}{(1-\mu)\varepsilon_\mathrm{sc}}\right) \eta (1-\mu)}.    
\end{equation}
When $R=R_0$, we have $t=t_\mathrm{acc}$ and $\xi=\xi_\mathrm{acc}\approx 5\xi_\mathrm{load}$, therefore the above expression simplifies to
\begin{equation}\label{t_hat_pairs_R0}
    \delta \hat{t}_{\rm eff}(\varepsilon_\mathrm{sc},\mu,t) = \frac{1}{2} \left(\frac{e^{(1+\rho_\mathrm{acc})}}{1+\rho_\mathrm{acc}}+\frac{e^{(1-\rho_\mathrm{acc})}}{1-\rho_\mathrm{acc}}-\frac{2 e^{-5\xi_\mathrm{load}}}{1-\rho_\mathrm{acc}^2}\right),
\end{equation}
which is the expression that we use in our model.

The optical depth due to the Thompson scattering is simply 
\begin{equation}
    \tau_{\rm \gamma e}^{\pm} ({\varepsilon},t) \sim
        \frac{\hat{A}}{2R_{0}} \sigma_{\rm KN}({\varepsilon}) (e^{\xi(R_0,t)/\xi_{\rm load}}+e^{-\xi(R_0,t)/\xi_{\rm load}})\sim \frac{\hat{A}}{2R_{0}} \sigma_{\rm KN}({\varepsilon}) (e^{\xi_\mathrm{acc}/\xi_{\rm load}}+e^{-\xi_\mathrm{acc}/\xi_{\rm load}})\approx \frac{74\hat{A}}{R_{0}} \sigma_{\rm KN}({\varepsilon})\approx 1.5\,A_\mathrm{\star,2}R_{0,15}^{-1}\frac{\sigma_\mathrm{KN}(\varepsilon)}{\sigma_\mathrm{T}},
\end{equation}
where the last three expressions assume that $\xi=\xi_\mathrm{acc}$.

\begin{figure*}
  \sidecaption
  \includegraphics[width=12cm]{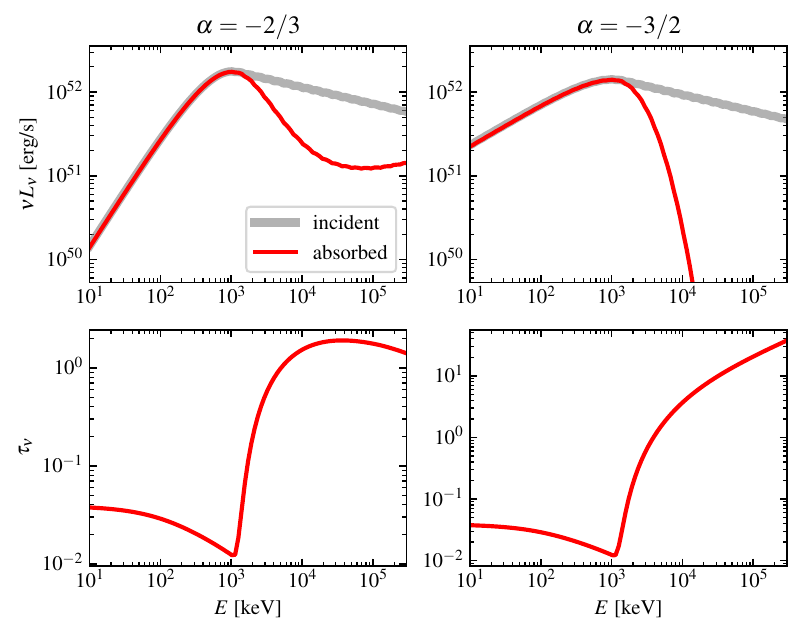}
  \caption{Example incident and absorbed spectra and corresponding optical depth for incident photons. Incident spectra (grey solid lines) are characterised by $\beta=-2.2$, $E_\mathrm{p}=1$ MeV, $L=10^{53}$ erg/s, and a low-energy photon index $\alpha=-2/3$ (left-hand panels, representative of synchrotron in marginally fast cooling) or $\alpha=-3/2$ (right-hand panels, for synchrotron in fast cooling). The absorbed spectra, assuming $t=1$ s (which implies $R_0=7.9\times 10^{15}\,\mathrm{cm}$) and $A_\star = 10^3\,\mathrm{cm^{-3}}$, are shown with red lines in the top panels. The corresponding optical depth, which includes both Thomson scattering and $\gamma-\gamma$ absorption, is shown in the bottom panels. For $\alpha=-2/3$, the absorption feature is saddle-shaped, while for $\alpha=-3/2$ it produces a spectral cut-off.\\[22pt]}
  \label{fig:example_abs_spectra}
\end{figure*}

Figure \ref{fig:example_abs_spectra} shows example incident and absorbed spectra calculated with our model for a particular choice of model parameters, highlighting the effect of assuming two different low-energy photon indices for the incident spectrum. The bottom panels show the total optical depth, which is dominated by Thomson scattering below 1 MeV and by $\gamma-\gamma$ absorption above. When the low-energy photon index is harder than -1, as expected in a marginally-fast-cooling synchrotron scenario \citep{Oganesyan2017} the absorption feature has a `saddle' shape; if the incident spectrum is softer, as expected in the usual fast-cooling synchrotron scenario, then the absorption induces a spectral cut-off.

\section{GRB 190114C fit results}

\begin{table}[t]
\centering
\setlength{\tabcolsep}{10pt} 
\renewcommand{\arraystretch}{1.2} 
\caption{Best fit parameters for the Band and absorbed Band models.}
\begin{tabular}{lc}
\hline\hline
Parameter & Value \\
\hline
\multicolumn{2}{c}{\textbf{Band model}} \\
\hline
$\alpha$ & $-0.558 \,\pm\, 0.009$ \\
$\beta$  & $-2.81 \,\pm\, 0.03$ \\
$E_{\rm p}/(1+z)$ [$\mathrm{keV}$] & $709^{+16}_{-13}$ \\
$\mathrm{norm}$ [$\mathrm{ph\;cm^{-2}\;s^{-1}\;keV^{-1}}$] & $0.6105^{+0.005}_{-0.006}$ \\
\hline
Stat/dof & $803/350$ \\
\hline
\multicolumn{2}{c}{\textbf{Absorbed Band model$^\mathrm{a}$}} \\
\hline
$\alpha$ & $-0.59 \,\pm\, 0.01$ \\
$\beta$  & $-2.03^{+0.03}_{-0.08}$ \\
$\log_{10}(E_{\rm p}/\mathrm{keV})$ & $3.01 \,\pm\, 0.01$ \\
$\log_{10}(L/\mathrm{erg \, s^{-1}})$ & $53.55^{+0.06}_{-0.12}$ \\
$\log_{10}(\hat{A}_\star/\mathrm{cm^{-3}})$ & $3.98^{+0.11}_{-0.22}$ \\
$\log_{10}(R_0/\mathrm{cm})^\dagger$ & $16.40^{+0.03}_{-0.05}$ \\
\hline
Stat/dof & $656/349$ \\
\hline
\end{tabular}
\\~
\tablefoottext{$^\mathrm{a}$}{The table reports the statistical error on the parameters, but the uncertainty is dominated by the systematic that stems from fixing the $\xi_\mathrm{acc}$ and $t_0$ parameters, see the main text.}
\tablefoottext{$\dagger$}{Derived parameter, computed through Eq.\ \ref{eq:R0}.}
\label{fit_results}
\end{table}

We report in Table \ref{fit_results} the results of fitting two spectral models to the \textit{Fermi} observations of GRB~1900114C during the first 4.15~s after the \textit{Fermi}/GBM trigger. The models are either Band, or a model where the incident spectrum is still described by the Band function, but the effect of our absorption process is taken into account. More details are given in the main text.  

\section{Impact of the CSM on the afterglow and consequences for GRB190114C}
\label{app:denseafterglow}

The possible presence of a dense CSM around the progenitor would affect the dynamics of the blast wave that arises from the interaction of the GRB ejecta with the external medium. The CSM rest mass implied by our parameters for GRB190114C is at least $M_\mathrm{CSM}\sim 0.1\,\hat A_{\star,3.5}R_{0,16}\,\mathrm{M_\odot}$: after crossing it, a blast wave with a total isotropic-equivalent energy $E_\mathrm{ej}$ would be limited to expanding at an average Lorentz factor $\Gamma_\mathrm{max}\sim\sqrt{E_\mathrm{ej}/M_\mathrm{CSM}c^2}\approx 30 E_\mathrm{ej,55}^{1/2}\hat A_{\star,3.5}^{-1/2}R_{0,16}^{-1/2}$. The external shock would start when the ejecta reach the `gap' radius $R_\mathrm{gap}\sim R_0/3$ where the pre-acceleration Lorentz factor equals the ejecta Lorentz factor \citep{Beloborodov2002}. From that radius on, during the initial phase of the deceleration---when the reverse shock has not crossed yet the ejecta---the Lorentz factor of the external shock downstream is approximately constant and equal to $\Gamma_\mathrm{CSM}\sim 30 L_\mathrm{ej,55}^{1/4}\hat A_{\star,3.5}^{-1/4}$ \citep{Beloborodov2014}, where $L_\mathrm{ej}$ is the isotropic-equivalent kinetic power of the ejecta. During this phase, the forward shock traverses a region with a rapidly varying pair-loading. Upscattering of prompt emission photons by pairs in the shock downstream might be the main mechanism producing the early GeV emission seen by LAT in this GRB \citep[][]{Beloborodov2014,Hascoet2015}. In the observer frame, the shock traverses the CSM over a time $\Delta t_\mathrm{CSM}\sim (1+z)R_\mathrm{0}/\Gamma_\mathrm{CSM}^2 c \approx 520 R_{0,16}\Gamma_\mathrm{CSM,1.5}^{-2}\,\mathrm{s}$. The TeV emission detected by MAGIC \citep{MAGIC1-2019} is therefore mostly produced during this phase. A relevant question, therefore, is whether the very-high-energy (VHE) photons can escape the shock downstream without undergoing $\gamma-\gamma$ annihilation.

In the fluid comoving frame, a VHE photon has an energy $E_\mathrm{VHE}^\prime = (1+z)E_\mathrm{VHE}/\Gamma_\mathrm{CSM}$, and it preferentially annihilates with target photons of observed energy $E_\mathrm{target}=m_\mathrm{e}^2c^4\Gamma_\mathrm{CSM}/(1+z)E_\mathrm{VHE}^\prime=m_\mathrm{e}^2c^4\Gamma_\mathrm{CSM}^2/(1+z)^2E_\mathrm{VHE}$. The number density of target photons in the comoving frame is $n^\prime_\gamma(E^\prime_\mathrm{target})=L_{E}(E_\mathrm{target})/4\pi R_0^2\Gamma_\mathrm{CSM} c=(1+z)^2E_\mathrm{target}L_{E}(E_\mathrm{target})E_\mathrm{VHE}/4\pi R_0^2\Gamma_\mathrm{CSM}^3 m_\mathrm{e}^2 c^5$. The optical depth to $\gamma-\gamma$ annihilation can be estimated as $\tau_{\gamma\gamma}(E_\mathrm{VHE})\sim \sigma_{\gamma\gamma}n^\prime_\gamma(E^\prime_\mathrm{target})\Delta R^\prime$, assuming a cross section $\sigma_{\gamma\gamma}\sim \sigma_\mathrm{T}/5$ and a shock downstream width $\Delta R^\prime\sim c t^\prime_\mathrm{cool}/3$, where $t^\prime_\mathrm{cool}$ is the radiative cooling time of relativistic leptons in the shock downstream whose energy is at least $E^\prime_\mathrm{VHE}$. This reflects the fact that VHE photons are produced close to the shock, by the most energetic leptons. Assuming the cooling to be dominated by inverse Compton scattering (as justified by the comparable synchrotron and inverse Compton luminosities; \citealt{MAGIC2-2019}), we have $t^\prime_\mathrm{cool}=E_\mathrm{VHE}^\prime/P^\prime_\mathrm{IC}\sim 4\pi R_0^2\Gamma_\mathrm{CSM}^3m_\mathrm{e}^2c^4/\sigma_\mathrm{T} L (1+z)E_\mathrm{VHE}$, where $P_\mathrm{IC}^\prime$ is the inverse Compton power of a single lepton with energy $E^\prime_\mathrm{VHE}$ and $L$ here is the luminosity in photons below the Klein-Nishina threshold for the lepton considered. After some algebra, this leads to
\begin{equation}
    \tau_{\gamma\gamma}(E_\mathrm{VHE})\sim  \frac{(1+z)E_\mathrm{target}L_{E}(E_\mathrm{target})}{15 L}\approx 0.1 \frac{E_\mathrm{target}L_\mathrm{E}(E_\mathrm{target})}{L}.
\end{equation}
Since the spectrum of the GRB190114C afterglow was rather flat at the time of the TeV emission \citep{MAGIC2-2019},  $E_\mathrm{target}L_\mathrm{E}(E_\mathrm{target})\sim L$ and hence the shock downstream was transparent to TeV photons.

When the blast wave shock encounters the large density contrast between the CSM ($\hat{A}_\star \sim 10^3-10^4\,\mathrm{cm^{-3}}$) and the `standard' progenitor wind ($\hat{A}_\star \sim 1\,\mathrm{cm^{-3}}$; \citealt{Panaitescu2002,Beniamini2015,Tiwari2025}), it can undergo a re-acceleration phase (e.g.\ \citealt{Perna2002}, but we already noted that $\Gamma_\mathrm{CSM}\sim \Gamma_\mathrm{max}$, therefore the acceleration is only very limited). After that phase, it coasts until it has swept a rest mass comparableto that of the CSM, which happens at a `re-deceleration' radius $R_\mathrm{re-dec}\sim 2\times 10^{18} E_\mathrm{ej,55}\Gamma_\mathrm{CSM,1.5}^{-2} A_{\star,0}^{-1}\,\mathrm{cm}$. In the observer frame, this corresponds to $t_\mathrm{re-dec}\sim (1+z)R_\mathrm{re-dec}/\Gamma_\mathrm{max}^2c\approx 1.2  E_\mathrm{ej,55}\Gamma_\mathrm{CSM,1.5}^{-4} A_{\star,0}^{-1}\,\mathrm{d}$. Intriguingly, at a similar time, the optical light curves show a slope change \citep{MAGIC2-2019}.

Clearly, whether the overall emission is compatible with our scenario remains an open question.

\end{appendix}

\end{document}